\documentclass[aps,showpacs,prd,tightenlines,twocolumn,10pt,nofootinbib]{revtex4}

\usepackage[latin1]{inputenc}
\usepackage{bbm}

\newcommand{\be}{\begin{eqnarray}} % only untightened
\newcommand{\ee}{\end{eqnarray}}
\newcommand{\nn}{~\nonumber \\}

\newcommand{\titel}[1]{~\\}

\usepackage{color}

\begin{document}

%%%%%%%%%%%%%%%%%%%%%%%%%%%%%%%%%%%%%%%%%%%%%%%%%%%%%%%%%%%%%

\title{
Slowly decaying classical fields, unitarity, and gauge invariance 
}

\author{Dennis D. Dietrich}
\affiliation{The Niels Bohr Institute, Copenhagen, Denmark}

\date{December 1, 2005}

%%%%%%%%%%%%%%%%%%%%%%%%%%%%%%%%%%%%%%%%%%%%%%%%%%%%%%%%%%%%%

\begin{abstract}

In classical external gauge fields that fall off less fast than the inverse 
of the evolution parameter (time) of the system the implementability of a 
unitary perturbative scattering operator ($S$-matrix) is not 
guaranteed, although the field
goes to zero. The importance of this point is exposed for the counter-example 
of low-dimensionally expanding systems.
The issues of gauge invariance and of the interpretation of the evolution at 
intermediate times are also intricately linked to that point.


\end{abstract}

%%%%%%%%%%%%%%%%%%%%%%%%%%%%%%%%%%%%%%%%%%%%%%%%%%%%%%%%%%%%%%

\maketitle

%%%%%%%%%%%%%%%%%%%%%%%%%%%%%%%%%%%%%%%%%%%%%%%%%%%%%%%%%%%%%

\section{Introduction}

This note deals with questions appearing frequently when considering processes
in classical fields. The principal question is whether there exists a
well-defined description for a generic scattering process in the presence of a
given background. After clarifying this point at least for asymptotically
early or late times, one may ask if information about the system 
can be extracted also at intermediate times. Here, the required criteria are exhibited and analysed for 
slowly decaying fields. In the course of the investigation this 
will also necessitate a discussion of gauge invariance. 

A classical field emerges as the expectation value of a quantised bosonic 
field of the underlying quantum field
theory. In situations where the occupation number of the bosonic field modes
is very large, the commutator of the bosonic field creation and 
annihilation operators is subleading. The dynamics of the
classical mean field capture a major part of the dynamics of the physical
system. Quantum effects are parametrically smaller. In abelian field theories 
at weak coupling, the leading quantum effects are due to fermions and
antifermions. In non-abelian field theories, quantum fluctuations of the 
bosonic field are parametrically equally favoured \cite{dddpert}. The currents 
induced by
the movement of these quanta in the classical field modify the current term
in the equations of motion for the classical field. Frequently this quantum
effect, the so-called back-reaction, is neglected in what is known as the external 
field approximation. While quantum effects are regarded as parametrically 
suppressed in the systems considered, they might still fundamentally alter
the systems' description beyond a mere correction to observables.

The present article addresses this point for the characterisation of a
system with a slowly decaying external field in terms of free particles. 
In the present sense, a slowly decreasing field means a gauge field which 
does decay to zero
but not faster than the inverse of the evolution parameter (time). In many 
cases, a change of gauge can lead to a faster decay. Here, however, we will
consider situations where this is not possible.
For example, one can imagine a volume expanding in one spatial
direction with a given constant total energy content in the classical field. 
Then, the corresponding energy density will decrease inversely with time,
and the components of the field tensor like the inverse of the square root of 
time. This is also the fastest possible decay for the largest component of the 
gauge field in any gauge.
Note that here, in the spirit of perturbation theory, the 
situation for asymptotically {\it free} particles is studied. For other 
asymptotic bases, the 
picture can be different. Actually, a change of basis can often cure
some of the problems one encounters as in the case of the slow spatial
decay (long-range interaction) of the field in Coulomb scattering.  

Beyond the external field approximation, the fall-off of the fields is 
accelerated by processes subsumed under back-reactions. For instance, 
accelerating or even producing particles by vacuum polarisation requires energy 
which must come from the classical field. Naturally, one can also imagine
situations where energy is transferred from the quantum sector to the classical 
field but in the present setting, where the classical field is to carry the
bulk of the energy initially, this contribution is relatively suppressed. As the 
incorporation of the back-reaction is usually a formidable task, one tends to 
work within the framework of the external field approximation. Then, however, 
one must also pay attention to potential fundamental problems \cite{aw} 
such as the non-unitarity of the theory.

In order to shed more light on these points, section \ref{unitarity} 
addresses the issue of unitarity in slowly decaying external fields. It is 
investigated whether a unitary implementation of the scattering operator for
fermions and antifermions exists in the presence of a slowly decreasing
field. (Although the implementation of the scattering operator in Fock space
is an operator in a Hilbert space, it will be called by its traditional name 
$S$-matrix, in what follows.) As shall be discussed below, this depends 
on whether the number of particles produced from the vacuum in the presence 
of the field is finite or not \cite{szpak,ss}. 
The requirement that the field should decay is a much weaker condition than the
known set of sufficient conditions \cite{t} guaranteeing a finite number of
produced particles. In fact, fields that decay slowly do not 
satisfy such conditions. However, the {\it minimal} sufficient conditions are 
not known. Thereby, on the one hand, the situation is unclear in many physically
interesting systems, and, on the other, has to be investigated separately for 
single or small classes of problems. 
The set of backgrounds which allow determination of the free 
particle content at any intermediate time is even more restricted than the set 
of backgrounds which permit a unitary perturbative $S$-matrix and does not
contain slowly decaying fields \cite{r,s,t}. In section \ref{moller} the
existence of the M{\o}ller wave operators in slowly decaying fields is
investigated 
in one spatial and one temporal dimension. In section \ref{us} the existence
of a unitary $S$-matrix is studied by explicitly calculating the expectation
value of the number of massive fermions produced in the presence of a slowly
decreasing field for three spatial and one temporal dimension. 
Calculational details are given in appendix \ref{appendix}.

Section \ref{gauge} is concerned with the issue of gauge invariance. 
The description of 
the production of free fermion-antifermion-pairs by vacuum polarisation in the
presence of classical fields with identified particle and antiparticle
momenta per se depends on the chosen gauge. As explained below, a
gauge independent interpretation can be given to the results.
However, for slowly decaying field configurations this is more
complicated to achieve.

Section \ref{conclusion} summarises the results.

%%%%%%%%%%%%%%%%%%%%%%%%%%%%%%%%%%%%%%%%%%%%%%%%%%%%%%%%%%%%%%%%%%%%%%%%%%%%%

\section{Unitarity\label{unitarity}}

A physical system can be described by the notion of in- and out-going {\it
asymptotic} particles if the scattering operator $s$ acting on the Hilbert
space $h$, \mbox{$s:h\rightarrow h$}, has a unitary implementation 
in the form of the operator $S$ (S-matrix)
in the Fock space $F$ constructed from the
corresponding asymptotic states: \mbox{$S:F\rightarrow F$}.
When talking about processes with in- and out-going {\it free} particles
one must check whether the unitary (perturbative) $S$-matrix for {\it free} 
states exists in the case under consideration.

A more stringent requirement on the relative system
than the existence of the unitary $S$-matrix is the implementability of unitary
evolution at any intermediate time. For fermions, this means 
that the time-development operator, $U(x_0,y_0)$, 
describes the system's evolution based on
the initial vacuum states and that at any, in general not asymptotically large
or small, time the particle content of the system is known.
The operator $U(x_0,y_0)$ evolves the wave-function solutions $\psi$ of the 
Dirac equation 
\be
i\partial_t\psi(t)=H(t)\psi(t),
\ee
according to:
\be 
\psi(x_0)=U(x_0,y_0)\psi(y_0).
\ee
$H(t)$ is the Dirac Hamiltonian.
The set of backgrounds for which this is actually possible is much 
too restricted \cite{r,t,s} to include slowly decaying fields.
(The
set of sufficient conditions given in \cite{r} admits only an $A_0$
component on a compact support. Thereby no magnetic fields are included.) 
Therefore,
projections onto asymptotic states at an intermediate time do not yield the 
number of particles in the system at that time for the situations
addressed here. Frequently, it is impossible even to define a particle in 
the presence of an external field for non-asymptotic times \cite{Fierz:1979ys}.

The conditions for the existence of the M{\o}ller wave operators are less 
stringent. They are defined as limits of products of full evolution operators 
$U$ with free evolution operators $U_0$ for infinitely early or late times:
\be
W_\pm(x_0):=\lim_{y_0\rightarrow\pm\infty}U(x_0,y_0)U_0(y_0,x_0).
\ee
$U_0$ is the evolution operator for the free Dirac equation:
\be
i\partial_t\psi_0(t)=H_0\psi_0(t).
\ee

If the M{\o}ller wave operators exist, the single-particle scattering operator 
exists as well and can be expressed with the help of the former:
\be
s=W_+^*W_-.
\ee

If it exists, $s$ describes the infinite initial and final time limit of the 
evolution. It contains the information on how the initial 
state evolves from infinitely early times to infinitely late times. 
For time-dependent Hamiltonians the M{\o}ller operators and $s$ depend on the
time $x_0$. The time-dependence reduces to a pure phase factor for the matrix
elements $\langle m|S|n\rangle$ (see below). This phase drops out of the
absolute squares of the matrix elements, which are the observable
quantities. Therefore, the time-dependence of the relative objects is not
denoted.

The criterion for when a unitary $S$-matrix can be constructed is 
formulated most conveniently after carrying out a decomposition. 
The projectors $P_{\pm}$ serve to project onto the positive and negative
energy sectors of the Hilbert space $h$: $P_\pm h=h_\pm$. With their 
help, the scattering operator $s$ and thereby its implementation $S$ on the 
Fock space $F$ can be decomposed according to 
\be
S_{\pm\pm}:=P_\pm SP_\pm,
\ee
where the first and second sign on each side of the equality are linked.
Hereafter, the upper or the lower signs have to be chosen in all
corresponding expressions.
The Shale-Stinespring criterion \cite{ss} now says that a unitary $S$-matrix 
exists iff $S_{\pm\mp}$
are Hilbert-Schmidt operators, i.e.
 have a finite Hilbert-Schmidt norm:
\be
\sum_n\langle n|S_{\pm\mp}^\dagger S_{\pm\mp}|n\rangle\in\mathbbm{R}^+,
\label{ss}
\ee
where the $|n\rangle$ form a complete set of states.
Interestingly, after inserting a complete set of states between the two
operators in the previous expression one sees that the criterion has a 
direct physical interpretation because:
\be
\sum_{m,n}
|\langle m|S_{\pm\mp}|n\rangle|^2
=
\sum_m N_m,
\label{interpretation}
\ee
where $N_m$ is the number of particles/antiparticles (upper/lower signs)
in mode $m$ produced from the vacuum \cite{szpak}; that is only if the
initial state is a vacuum. In other words, if a unitary $S$-matrix is to
exist, the number of (anti)particles produced from the vacuum must be finite 
in every single mode $m$ and in total.

In fact, in order to describe the system's particle content at any 
intermediate point in time, the evolution operator $U$ must also be a 
Hilbert-Schmidt operator. This, however, constrains the possible forms of the 
field much more severely (see above) than in the case of the M{\o}ller wave operators 
and the scattering operator. 

A final important remark is due at this point: In systems in which the
{\it perturbative} $S$-matrix (i.e. the one formulated for {\it free} 
states) does not exist, an $S$-matrix based on different {\it asymptotic}
states may well exist. A prominent example is the scattering in
unscreened Coulomb fields. When formulated with free states it is 
plagued by a logarithmic phase as a function of the distance from the source 
of the field. Thus it grows without bound as the distance increases. There, 
this problem can be avoided by
formulating scattering using the corresponding continuum Sommerfeld-Maue 
wave-functions which leads to Dollard modified wave operators.

% % % % % % % % % % % % % % % % % % % % % % % % % % % % % % % % % % % % % % 

\subsection{Existence of M{\o}ller operators\label{moller}}

First, let us look at the question of whether the M{\o}ller operators
exist in one temporal and one spatial dimension for massless fermions in a
slowly decaying abelian field. In such a field the full Hamilton 
operator reads:
\be
H=-A_0(t,z)-\gamma^0\gamma^3[i\partial_3+A_3(t,z)]
\ee
and the free Hamilton operator is given by:
\be
H_0=-i\gamma^0\gamma^3\partial_3.
\ee
Then, what must be examined in order to establish the existence of the
M{\o}ller operators are the limits $y_0\rightarrow\pm\infty$ of:
\be
U(x_0,y_0)U_0(y_0,x_0)
=
\rho^+e^{iu^+}+\rho^-e^{iu^-},
\ee
with
\be
u^+
&:=&
a_+(x_+,x_-)-a_+[x_+,x_-+\sqrt{2}(y_0-x_0)]
\nn
u^-
&:=&
a_-(x_+,x_-)-a_-[x_++\sqrt{2}(y_0-x_0),x_-]
\ee
and the projectors \mbox{$2\rho^\pm:=1\pm\gamma^0\gamma^3$} as well as the 
antiderivatives:
\be
a_\pm(x_+,x_-):=\int dx_\mp A_\pm(x_+,x_-),
\ee
where the light-cone coordinates are defined as
\mbox{$\sqrt{2}v_\pm:=v_0\pm v_3,~v\in\{x;A\}$}.
As the exponential is a continuous function, the question is whether the limits
exist for its exponent. Hence, the crucial limits are:
\be
\lim_{x\pm\rightarrow c\infty}a_\mp(x_+,x_-)\stackrel{?}{\in}\mathbbm{R}
\label{limits}
\ee
with \mbox{$c\in\{-1;+1\}$}. For example, the limits towards $+\infty$
($c=+1$) do not exists in the presence of purely time-dependent or 
longitudinally boost-invariant fields, which decay slowly. Therefore, the 
M{\o}ller wave operator $W_+$ does not exist.

%  %  %  %  %  %  %  %  %  %  %  %  %  %  %  %  %  %  %  %  %  %  %  %  %  %

\subsection{Unitarity of the $S$-matrix\label{us}}

In addition to the results for the M{\o}ller operators, we now calculate 
explicitly the number of particles produced in the presence of a slowly 
decaying field. As mentioned above, the theory possesses a unitary formulation 
only if this number is finite.
To this end, the expectation value of the number of produced pairs is to be 
calculated in the special field:
\be
A_\mu=A_\mu(t)=a~g_{3\mu}~\theta(t)/\sqrt{t},
\label{unregularised}
\ee
This expectation value can be written as the double phase-space
integral,
\be
\langle n\rangle
=
\int
\frac{d^3p}{2(2\pi)^3\omega_p}
\frac{d^3q}{2(2\pi)^3\omega_q}|M_{q,p}|^2,
\label{expectation}
\ee
of the absolute square of the amplitude:
\be
M_{q,p}
=
\lim_{x_0\rightarrow +\infty\atop y_0\rightarrow -\infty}
\int
d^3xd^3y
\phi_q^\dagger(x)G_R(x,y)\gamma^0\psi_p(y),
\label{amplitude}
\ee
which is equal to the overlap of an incoming free antiparticle $\psi_p(y)$ 
with an outgoing free particle $\phi_q(x)$ after propagation through the 
field \cite{ev,dddst1,dddst2}. $G_R(x,y)$ represents the retarded propagator 
to all
orders in the field and \mbox{$\omega_p:=\sqrt{|\vec p|^2+m^2}$}. In 
Eq.~(\ref{expectation}) the summation over all discrete degrees of freedom 
is implicit.

In order to make the link, we note that
the double phase-space integral in Eq.~(\ref{expectation}) corresponds to the 
summation over the complete sets of states in Eq.~(\ref{interpretation}) and
that the amplitude (\ref{amplitude}) is equivalent to the matrix element in 
 Eq.~(\ref{interpretation}). The homogeneous solution $G_H(x,y)$ of the Dirac
equation in Eq.~(7) of \cite{dddst2} is equivalent to the evolution operator 
$U(x_0,y_0)$ if the former is taken at $\vec x=\vec y$ and the derivative 
operators acting on $\vec y$ instead of $\vec x$. Based on the homogeneous 
solution $G_H(x,y)$, the retarded propagator $G_R(x,y)$ appearing in the
amplitude (\ref{amplitude}) can be expressed as \cite{dddst1,dddst2}:
\be
iG_R(x,y)\gamma^0
=
+G_H(x,y)\delta^{(3)}(\vec x-\vec y)\theta(x_0-y_0).
\ee
The late-time behaviour of the gauge field is most important for
particles with long wavelengths. Therefore, the behaviour of the expectation
value for small energies has to be investigated. As can be seen by putting
Eqs.~(\ref{B}), (\ref{N0}), (\ref{N1}), and (\ref{N2}) into Eq.~(\ref{MTF}) 
and the latter into Eq.~(\ref{EV}), the number of pairs produced in the
presence of the field (\ref{unregularised}) at low energies is infinite [up
to finite corrections of the order \mbox{${\cal O}(\omega^2/a^2)$}] because of the
divergent expressions (\ref{C0}), (\ref{C1}), and (\ref{C2}) as well as the
pole in Eq.~(\ref{B}). 
For this reason, there exists no unitary perturbative $S$-matrix in this 
system. This is consistent with the observation that the M{\o}ller wave 
operators are ill-defined.
Actually, in this context the existence of the M{\o}ller operators can be 
seen as a necessary condition.

In order to understand how this result emerges and how to overcome the
inconsistency, let us investigate the same quantity in the regularised
field:
\be
A_\mu=A_\mu(t)=a~g_{3\mu}~\theta(t)~e^{-\beta\sqrt{t}}/\sqrt{t}.
\label{regularised}
\ee
In this case, all M{\o}ller wave operators exist because the limits 
(\ref{limits}) exist.
For this field one finds by introducing Eqs.~(\ref{b}) and 
(\ref{n}) into (\ref{MTF}) and subsequently into (\ref{EV}) that
\be
\frac{(2\pi)^3}{V}\frac{d\langle n\rangle}{d^3p}
=
8\frac{{m_T}^2}{\omega^2}
\left[4\frac{a}{\beta}-\sin\left(2\frac{a}{\beta}\right)\right]^2
+
{\cal O}\left(\frac{\omega^2}{\beta^4}\right),
\nonumber
\ee
which is finite for any non-zero value of the parameter $\beta$. The
problem, that a unitary $S$-matrix cannot be constructed due to a violation of 
the criterion (\ref{ss}) at small energies, is absent. Further, no meaningful
limit \mbox{$\beta\rightarrow 0$} exists because the first term
in the square brackets diverges and the second is non-analytic. Subleading
terms, which become important for decreasing $\beta$ at fixed $\omega$,
finally lead to the small energy limit of Eq.~(\ref{C0}), which is infinite.
Thus, we again see that no meaning can be given to the unregularised
situation defined by the field (\ref{unregularised}).

This calculation can also be carried out for boost-invariant field 
configurations, and the principal outcome will be the same.
While the longitudinal length $L$ here factors out as part
of the three-volume $V$, the result there becomes independent of the rapidity 
$y$.

The regularisation of the result by exponentially decaying functions
leads to the introduction of the decay parameter $\beta$ on which the final
result now depends. One cannot eliminate it right away without
encountering the same problems as before. One must either give a physical
meaning to the parameter, or one must calculate the effect of the back-reaction in
this regularised version. After including the back-reaction in an
appropriate way, the limit
$\beta\rightarrow 0$ of the expectation value should be finite because 
the mathematical inconsistency is ultimately not a problem of physics
but of computational ability, analytical as well as numerical.

Yet another word of caution is due in connection with these so-called 
asymptotic switching constructions. For example, while an {\it eternally}
constant field tensor does not produce any particles \cite{b}, particle
production is described when temporal damping is included and the limit
of no damping is taken at the end \cite{s,switch,schwinger}. In general, 
the introduction of the switching function can fundamentally alter the 
physical content of the 
calculation. Another time-independent example that requires regularisation is
the Coulomb field, e.g. leading to its Yukawa screened form.

One more source for the artificial alteration of the result by the above
manipulations occurs if the decay parameter is assigned a physically motivated
value which turns out to lead to rapid decrease. Then, the
characteristics of the unregularised field can be hidden and the unduly fast
decay of the gauge field can lead to an unduly large
field tensor through its derivative. With a cut-off at $t_{\mathrm{max}}$, 
this is even more obvious because it induces a $\delta$-peak in the field
tensor:
\mbox{$
\vec E(t,\vec x)
=
\vec A(t_{\mathrm{max}},\vec x)\delta(t-t_{\mathrm{max}})
$}. 
For a slowly decaying function, this contribution cannot safely be left 
uncompensated and should be estimated. In addition,
such spikes contain all frequencies and lead to a distortion of the
spectrum. 
Note, however, that situations without magnetic field can be described
entirely by the $A_0$ component of the gauge field, if an adequate gauge is
chosen. Then the spike with its shortcomings is absent. This circumstance is 
reminiscent of the sufficient conditions for the unitary implementability of
the evolution operator $U$ (see section \ref{us}) \cite{r}.
Nevertheless, even for $\vec A=\vec 0$, an abrupt cutoff leads to differences
in observables like, for example, the number of produced particles.
Ultimately, these originate from the fact that with, respectively without the 
switching one finds oneself in different Fock spaces.

%%%%%%%%%%%%%%%%%%%%%%%%%%%%%%%%%%%%%%%%%%%%%%%%%%%%%%%%%%%%%%%%%%%%%%%%%%%%

\section{Gauge invariance\label{gauge}}

We start the discussion of gauge invariance for situations in which 
gauges can be found such that the
M{\o}ller wave operators exist. The gauges for which this is the case
for a given system are connected by transformations, $\omega(x)$, which 
approach unity sufficiently fast for early and late times. In all such
gauges, 
the amplitude (\ref{amplitude}) remains the same. This is true because the 
propagator $G_R(x,y)$ in Eq.~(\ref{expectation}) transforms under an arbitrary 
gauge transformation 
$\Omega(x)$ like
\be
G_R(x,y)\rightarrow\Omega^\dagger(x)G_R(x,y)\Omega(y).
\ee
The gauge transformations at all intermediate times $z_0$ with
\mbox{$x_0>z_0>y_0$} drop out. Due to the limits taken in
Eq.~(\ref{amplitude}), the amplitude $M_{q,p}$ does not change because
\be
\lim_{
\scriptsize
\begin{array}{c}x_0\rightarrow +\infty\\y_0\rightarrow -\infty\end{array}
}
\omega^\dagger(x)G_R(x,y)\omega(y)
=
\lim_{
\scriptsize
\begin{array}{c}x_0\rightarrow +\infty\\y_0\rightarrow -\infty\end{array}
}
G_R(x,y).
\nn
\ee
Now, one possibility is to say that the gauge invariance of the theory is
tantamount to invariance under transformations $\omega(x)$.
Alternatively, one can allow all gauge transformations
but define the asymptotic states as
\be
\tilde\psi_p(x):=\tilde\Omega(x)\psi_p(x),
\ee 
where $\tilde\Omega(x)$ is a gauge transformation that for very
early and very late times leads to the same pure gauge field as the
transformation $\Omega(x)$ \cite{dddst1,dddst2}. In principle 
\mbox{$\tilde\Omega(x)=\Omega(x)$} could be chosen right away, but this is not 
the only choice. It does not influence any observables, and one is often not 
in the position to uniquely identify $\Omega(x)$ for lack of 
a reference. The approach admitting all gauge transformations is more in the spirit of a gauge theory as it is
based on the fact that one cannot only describe an asymptotic fermion as an 
entirely free particle but, by consistently transforming the whole theory, 
also as one in a pure gauge field. This procedure is closely related to
including Wilson lines with the propagator. There, the freedom of choice is
linked to the selection of the path over which the link is to be evaluated.
At the level of the M{\o}ller operators, this corresponds to incorporating the 
pure gauge field induced by $\Omega(x)$ into the free Hamiltonian $H_0$. At
that level, the procedure is unique as opposed to the choice of
$\tilde\Omega(x)$ relative to $\Omega(x)$ above. However, the identification
of the variables conjugate to the position variables of the necessary two-point
function with free particle momenta is straight forward only in the first
approach using the constrained set of gauge transformations.

Now let us continue by turning to slowly decaying fields. 
As has been discussed in section \ref{unitarity}, a slowly decaying field
can prevent a well-defined description of the system. In that case a 
regularisation has to be carried out to have a meaningful theory.
Regularisations on the basis of the gauge field can be carried out in any
gauge whence they are not unique and, in general, hinder a gauge invariant
interpretation of the result. For this reason the regularisation must be 
carried out for gauge covariant objects like the field tensor because they
transform homogeneously. In a thus regularised theory it is again possible
to follow the steps explained in section \ref{unitarity} for obtaining a
gauge invariant interpretation of the result. This means for example that,
for a given regularisation carried out for the field tensor, any gauge that
leads to a fast decaying field should lead to the same result for
observables for free particles.

The most drastic regularisation is the cut-off. If applied to a slowly 
decaying gauge field it leads to significant artefacts 
and is therefore highly dangerous as discussed in section \ref{unitarity}. If 
used for the
field tensor, the gauge field is switched abruptly to a pure gauge. Although
the pure gauge field corresponds to a vanishing field tensor, it must decay
in time if a projection onto free states is to be used finally. For a
given situation, it is not clear that there exist gauges in which the
continuity of the gauge field at the cut-off and a sufficiently 
fast decay for large times can be achieved simultaneously. The continuity of
the gauge field at the cut-off is again required to avoid cut-off artefacts.
An induced spike in the field tensor leads to unnatural effects. But even
without the spike, discontinuities in the gauge field have an influence on
observables.

%%%%%%%%%%%%%%%%%%%%%%%%%%%%%%%%%%%%%%%%%%%%%%%%%%%%%%%%%%%%%%%%%%%%%%%%%%%%

\section{Conclusion\label{conclusion}}

The issues of unitarity and gauge invariance have been discussed for 
asymptotically free particles in slowly decaying backgrounds.
Here the emphasis is on 
fermions. Note, however, that in non-abelian theories bosonic quantum 
fluctuations are parametrically equally favoured as the fermions and must be 
taken into account in a complete treatment of the system. 

The projection of the wave function onto free states at non-asymptotic times 
does not yield the free
particle content at that time but can only be interpreted as a cut-off of
the field with all the usual caveats caused by the induced delta-spike in
the field tensor. 

The regularisation of a slowly decaying field may be an
absolute necessity because it does not allow for a unitary formulation of the
theory in the first place. In other words, it has to be given a faster decay
either by means of the incorporation of neglected phenomena or artificially.
Therefore, a 
gauge invariant interpretation of the result is more difficult than in 
situations with a fast decaying field because the regularisation has to be
carried out in a gauge invariant manner.

Hence, the conclusion must be that the description of the scattering of 
{\it free} particles in slowly decaying backgrounds is, in general, not a 
well-defined problem. Usually this is caused by the neglect of effects which 
are important for the modelling of the system. In the present case, their
inclusion would
ensure a sufficiently fast decrease of the classical field. Here, the
required effect that comes to mind first is the inclusion of the 
back-reaction. In the introduction we mentioned low-dimensional
expanding systems as examples of environments in which slowly decreasing
fields appear. In that case the small number of dimensions into which the
field expands may also be an approximation. Dropping this constraint 
can be a practical step which alleviates the difficulties by accelerating the 
gauge field's decay.

For practical purposes the inclusion of the back-reaction is a difficult
task. It is seldom possible to incorporate it into the
calculations exactly because, ultimately, this amounts to solving the full
theory. A standard loop expansion around the classical limit will, in general,
also be plagued by divergences in every order of the series.
Let this not be misunderstood; the classical calculation for the
bosonic field in itself is consistent even though it might not provide an 
accurate picture of the system. Additionally,
in a situation where a sizeable expectation
value for the bosonic field is present the expansion around zero 
field fails. The problems emerge when one has to go beyond 
the classical limit. Already at the prequantum level, 
concretely for classical Dirac 
particles in the background field, problems can arise in form of the 
non-existence of the M{\o}ller operators. Through the successive inclusion 
of quantum fluctuations the framework becomes yet more prone to
inconsistencies. One observes that the description within the usual
framework cannot be executed straight forwardly in situations which seem to
be most natural.
In other words, in situations in which the classical gauge field varies 
(decays) slowly, it provides an appropriate approximate description of the 
system. However, in this case, 
the standard procedure for the incorporation of quantum effects beyond the 
classical approximation is ill-defined. To the contrary, if the field varies 
(decays) rapidly the classical approximation is less good. Nevertheless, the
quantum corrections can be calculated within the standard approach.

In physics, other similar conceptual problems are known, for example within the
Thomas-Fermi method. The description of an almost homogeneous electron gas
by means of a gradient expansion around the homogeneous limit is flawed
already for arbitrarily small perturbations. This is the case although, in
this setting, the 
homogeneous approximation is very good. The conditions for the applicability
of the gradient expansion require low-amplitude, short-wavelength 
perturbations and forbid them at the same time \cite{hk}.

Finally, a loop expansion has to be amended by a self-consistency 
condition in order to become viable. However, commonly used iterative 
procedures which are adapted for numerical studies are frequently not 
transparent 
when it comes to understanding their physical content. Consequently, it is 
worth considering a fundamental reorganisation of the expansion scheme.

%%%%%%%%%%%%%%%%%%%%%%%%%%%%%%%%%%%%%%%%%%%%%%%%%%%%%%%%%%%%%%%%%%%%%%%%%%%%

\section*{Acknowledgments}

The author feels particularly indebted towards Andrew D.~Jackson for his
support during the course of the project; for inspiring discussions as well 
as the careful reading of and useful comments on the manuscript. 
Further helpful and informative discussions with Keijo Kajantie, Tuomas Lappi, 
Joachim Reinhardt, Dirk Rischke, Francesco Sannino, Kim Splittorff, 
Boris Tomasik, and especially Nikodem Szpak are acknowledged gratefully. 
Again thanks are due to Francesco Sannino and Nikodem Szpak for feedback on 
the manuscript.

%%%%%%%%%%%%%%%%%%%%%%%%%%%%%%%%%%%%%%%%%%%%%%%%%%%%%%%%%%%%%%%%%%%%%%%%%%%%

\appendix

\begin{widetext}

\section{Expectation value for the number of produced pairs\label{appendix}}

Making use of the implicit definition of the momentum transfer function 
${\cal T}$:
\be
G(x,y)
=
G^0(x-y)
+
\int d^4\xi d^4\eta G^0(x-\xi){\cal T}(\xi,\eta)G^0(\eta-y),
\ee
the amplitude (\ref{amplitude}) can be reexpressed so that the expectation
value (\ref{expectation}) in a purely time
dependent field becomes:
\be
\frac{4(2\pi)^3}{V}\frac{d\langle n\rangle}{d^3k}
=
\mathrm{tr}
\left\{
{\cal T}_R(+k_0,+\vec k;-k_0,+\vec k)
\frac{\gamma^0\omega-\vec\gamma\cdot\vec k-m}{\omega}
{\cal T}_A(-k_0,+\vec k;+k_0,+\vec k)
\frac{\gamma^0\omega+\vec\gamma\cdot\vec k+m}{\omega}
\right\}.
\label{EV}
\ee
Due to the relation \mbox{${\cal T}_A(p,q)=-{\cal T}_R^*(q,p)$} it does
suffice to give, for example, the retarded ${\cal T}$ in a field 
\mbox{$A_\mu(x)=g_{\mu3}~A_3(t)$}:
\be
{\cal T}_R
&=&
a\gamma^3{\cal T}^B
-
ia^2\sum_n
\gamma^3[{\cal T}_+^{(n)}\rho^++{\cal T}_-^{(n)}\rho^-]
[-i\gamma^0(\vec\gamma_\perp\cdot\vec k_\perp+m)]^n
\gamma^0\gamma^3
\label{MTF}
\ee
According to \cite{dddt} for the field 
\mbox{$A_\mu=g_{\mu3}~a~\theta(t)~t^{-\frac{1}{2}}$}
the coefficient functions ${\cal T}^B$ (Born) and ${\cal T}_\pm^{(n)}$ are
given by:
\be
{\cal T}^B=\sqrt{\frac{\pi}{\omega}}\frac{1+i}{2}
\label{B}
\ee
\be
{\cal T}_\pm^{(0)}
&=&
\int_0^\infty dx_0\int_0^{x_0}dy_0
{x_0}^{-\frac{1}{2}}{y_0}^{-\frac{1}{2}}
e^{+i\omega(x_0+y_0)}e^{\pm ik_3(x_0-y_0)}
e^{\pm2ia({x_0}^{\frac{1}{2}}-{y_0}^{\frac{1}{2}})}
=
\nn
&=&
4\int_0^1dc
[\pm2i(1-c)]^{-1}\frac{\partial}{\partial a}
\frac{1}{2}\sqrt{\frac{\pi}{-i\omega(1+c^2)\mp ik_3(1-c^2)}}
\times
\nn
&&\times
\exp\left\{\frac{[\mp ia(1-c)]^2}{-i\omega(1+c^2)\mp ik_3(1-c^2)}\right\}
\mathrm{erfc}\left[\frac{\mp ia(1-c)}{\sqrt{-i\omega(1+c^2)\mp
ik_3(1-c^2)}}\right]
=
\nn
&=&
\left(\frac{\partial}{\partial a}a^{-1}\right)
\left[
\int_0^1dc(1-c)^{-2}
+
{\cal O}\left(\frac{\omega^2}{a^2}\right)
\right]
\label{N0}
\ee
\be
\int_0^1dc(1-c)^{-2}
\rightarrow
\infty
\label{C0}
\ee
\be
{\cal T}_\pm^{(1)}
&=&
\int_0^\infty dx_0\int_0^{x_0}dy_0{x_0}^{-\frac{1}{2}}{y_0}^{-\frac{1}{2}}
e^{+i\omega(x_0+y_0)}e^{\pm ik_3(x_0+y_0)}
e^{\pm2ia({x_0}^{\frac{1}{2}}-{y_0}^{\frac{1}{2}})}
\int_{y_0}^{x_0}dz_0
e^{\mp2ik_3z_0}e^{\mp4ia{z_0}^{\frac{1}{2}}}
=
\nn
&=&
8\int_0^1dc_1\int_{c_1}^1dc_2c_2
[\pm2i(1+c_1-2c_2)]^{-3}
\frac{\partial^3}{\partial a^3}
\frac{1}{2}
\sqrt{\frac{\pi}{-i\omega(1+{c_1}^2)\mp ik_3(1+{c_1}^2-2{c_2}^2)}}
\times
\nn
&&\times
\exp
\left\{
\frac
{[\mp ia(1+c_1-2c_2)]^2}
{-i\omega(1+{c_1}^2)\mp ik_3(1+{c_1}^2-2{c_2}^2)}
\right\}
\mathrm{erfc}
\left[
\frac
{\mp ia(1+c_1-2c_2)}
{\sqrt{+i\omega(1+{c_1}^2)\pm ik_3(1+{c_1}^2-2{c_2}^2)}}
\right]
=
\nn
&=&
-\frac{1}{2}\left(\frac{\partial^3}{\partial a^3}a^{-1}\right)
\left[
\int_0^1dc_1\int_{c_1}^1dc_2c_2(1+c_1-2c_2)^{-4}
+
{\cal O}\left(\frac{\omega^2}{a^2}\right)
\right]
\label{N1}
\ee
\be
\int_0^1dc_1\int_{c_1}^1dc_2c_2(1+c_1-2c_2)^{-4}
\rightarrow
\infty
\label{C1}
\ee
\be
{\cal T}_\pm^{(2)}
&=&
\int_0^\infty dx_0\int_0^{x_0}dy_0{x_0}^{-\frac{1}{2}}{y_0}^{-\frac{1}{²}}
e^{+i\omega(x_0+y_0)}
e^{\pm ik_3(x_0-y_0)}
e^{\pm2ia({x_0}^{\frac{1}{2}}-{y_0}^{\frac{1}{2}})}
\times
\nn
&&\times
\int_{y_0}^{x_0}dz_0
e^{\mp2ik_3z_0}
e^{\mp4ia{z_0}^{\frac{1}{2}}}
\int_{y_0}^{z_0}dt_0
e^{\pm2ik_3t_0}
e^{\pm4ia{t_0}^{\frac{1}{2}}}
=
\nn
&=&
8\int_0^1dc_1\int_{c_1}^1dc_2c_2\int_{c_1}^{c_2}dc_3c_3
[\pm2i(1-c_1-2c_2+2c_3)]^{-5}
\frac{\partial^5}{\partial a^5}
\sqrt{\frac{i\pi}{\omega(1+{c_1}^2)\pm k_3(1-{c_1}^2-2{c_2}^2+2{c_3}^2)}}
\times
\nn
&&\times
\exp
\left\{
\frac
{[\mp2ia(1-c_1-2c_2+2c_3)]^2}
{-i\omega(1+{c_1}^2)\mp ik_3(1-{c_1}^2-2{c_2}^2+2{c_3}^2)}
\right\}
\mathrm{erfc}
\left[
\frac
{\mp2ia(1-c_1-2c_2+2c_3)}
{\sqrt{-i\omega(1+{c_1}^2)\mp ik_3(1-{c_1}^2-2{c_2}^2+2{c_3}^2)}}
\right]
=
\nn
&=&
\frac{1}{4}\left(\frac{\partial^5}{\partial a^5}a^{-1}\right)
\left[
\int_0^1dc_1\int_{c_1}^1dc_2c_2\int_{c_1}^{c_2}dc_3c_3(1-c_1-2c_2+2c_3)^{-6}
+
{\cal O}\left(\frac{\omega^2}{a^2}\right)
\right]
\label{N2}
\ee
\be
\int_0^1dc_1\int_{c_1}^1dc_2c_2\int_{c_1}^{c_2}dc_3c_3(1-c_1-2c_2+2c_3)^{-6}
\rightarrow
\infty
\label{C2}
\ee

%%%%%%%%%%%%%%%%%%%%%%%%%%%%%%%%%%%%%%%%%%%%%%%%%%%%%%%%%%%%%%%%%%%%%%%%%%%

According to \cite{dddt} one finds for the coefficient functions in the 
regularised field 
\mbox{$A_\mu=g_{\mu3}~a~\theta(t)~t^{-\frac{1}{2}}~e^{-\beta t^{\frac{1}{2}}}$}:
\be
{\cal T}^B
=
\sqrt{\frac{\pi}{-2i\omega}}\exp\left(\frac{\beta^2}{-8i\omega}\right)
\mathrm{erfc}\left(\frac{\beta}{\sqrt{-8i\omega}}\right)
=
-\frac{2}{\beta}+{\cal O}\left(\frac{\omega}{\beta^3}\right)
\label{b}
\ee
\be
{\cal T}_\pm^{(0)}
&=&
\int_0^\infty dx_0\int_0^{x_0}dy_0{x_0}^{-\frac{1}{2}}{y_0}^{-\frac{1}{2}}
e^{-\beta{x_0}^{\frac{1}{2}}}e^{-\beta{y_0}^{\frac{1}{2}}}
e^{+i\omega(x_0-y_0)}e^{\pm ik_3(x_0-y_0)}
e^{
\mp2i\frac{a}{\beta}
(
e^{-\beta{x_0}^{\frac{1}{2}}}
-
e^{-\beta{y_0}^{\frac{1}{2}}}
)
}
=
\nn
&=&
2\sum_{\nu=0}^\infty\sum_{\mu=0}^\infty\frac{1}{\nu!}\frac{1}{\mu!}
\left(\pm2i\frac{a}{\beta}\right)^\nu\left(\mp2i\frac{a}{\beta}\right)^\mu
\int_0^1dc\frac{1}{-[(\nu+1)+(\mu+1)c]}\frac{\partial}{\partial\beta}
\times
\nn
&&\times
\sqrt{\frac{i\pi}{\omega(1+c^2)\pm k_3(1-c^2)}}
\exp\left(
\frac{\{\beta[(\nu+1)+(\mu+1)c]/2\}^2}{-i[\omega(1+c^2)\pm k_3(1-c^2)]}
\right)
\mathrm{erfc}\left\{
\frac{\beta[(\nu+1)+(\mu+1)c]/2}{\sqrt{-i[\omega(1+c^2)\pm k_3(1-c^2)]}}
\right\}
=
\nn
&=&
-a^{-2}\left(e^{\pm2i\frac{a}{\beta}}-1\mp2i\frac{a}{\beta}\right)
+
{\cal O}\left(\frac{\omega}{\beta^3}\right)
\label{n}
\ee

\end{widetext}

%%%%%%%%%%%%%%%%%%%%%%%%%%%%%%%%%%%%%%%%%%%%%%%%%%%%%%%%%%%%%%%%%%%%%%%%%%%%%

\end{document}